# Differential equation of an aspherical lens and its solution

Ilia Evteev
Affiliation: Independent researcher
Email: iliadznt@yandex.ru

**Abstract**

A method for calculating the meridian profile of the aspherical surface of a plane-convex lens excluding spherical aberration, without calculating the coefficients of the series, by direct solving the compiled differential equation is proposed. The differential equation of an aspherical lens was compiled, found its numerical solution, and got the meridian profile $r(\varphi)$ of an aspherical lens. For long-focus lenses found conditions to simplify the differential equation.

## 1. Introduction

As is known, conventional lenses give spherical aberrations. In order to get rid of them, it is necessary to make the surface of the lens aspheric. The equation of the meridional section of an aspherical surface of revolution having an axis of symmetry looks like[1]-[4]

$$y^2 = A_1|x| + A_2 x^2 + A_3|x^3| + ... + A_k x^k \tag{1}$$

In most cases, optics uses aspherical surfaces - paraboloids, ellipsoids, hyperboloids, oblate spheroids obtained by rotating conical sections, which is given by equation (origin of the Cartesian coordinate system at the vertex of the meridian profile of surface and axis of symmetry combined with the *x*-axis) [4]

$$y^2 = 2rx + \left[e^2 - 1\right]x^2 \tag{2}$$

where $r$ - is the radius of curvature of the surface at its vertex, $e$ – eccentricity. Expression (2) contains one independent parameter $e^2$ that is used to correct aberrations. The expression for the deformed conic section is also often used [4]

$$y^2 = 2rx + \left[e^2 - 1\right]x^2 + \alpha x^3 + \beta x^4 \tag{3}$$

where $\alpha$ and $\beta$ - are constants.
There are also other forms of recording an equation of aspherical surface profile, for example

$$x = \left[(y^2 r)/(1+\sqrt{1-(1+e)y^2/r^2})\right] + \sum A_i y^{2i} + \sum B_i |y|^{2i-1} \tag{4}$$

where $r$ - is the radius of curvature of the surface at its vertex, e, $A_i$ $B_i$ - are constants that determine this particular type of surface [5]. For example coefficients $A_k$ can be determined through the values of the large and small semiaxes of the aspherical surface
Surfaces that differ little from the sphere are often described by the equation in the polar coordinate system

$$R = R_0 + A_1 \varphi^2 + A_2 \varphi^4 + A_3 \varphi^6 + ... \tag{5}$$

where $R_0$ - radius-vector of the vertex of the curve, and $\varphi$ is the polar angle of the profile point [5].

Thus, the standard way of eliminating aberrations of optical systems is to calculate or choose coefficients of the series $A_k$ before degrees of *x* or $\varphi$ in (1) or (5). And the accuracy of this method depends on the number of terms of the series 1. In this connection, there arises the problem of obtaining an aspherical profile of the lens, which would not depend on the number of refining terms of the series, but would represent a solution of some equation.

## 2. Equation building and results of solving

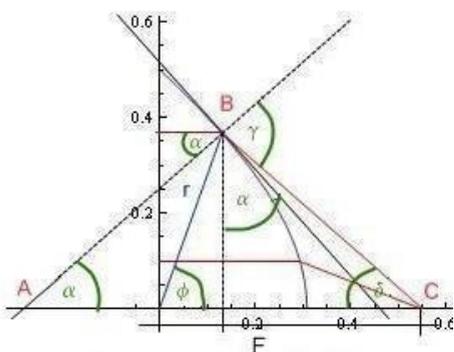

**Figure 1.** Ray paths in aspherical lens.

In this paper a method for finding the meridian profile of the aspherical surface of a plane-convex lens excluding spherical aberration, without calculating the coefficients of the series, by directly solving the compiled differential equation is proposed. So, the problem is formulated as follows: to compile the differential equation of an aspherical lens and solving it, get the meridian profile $r(\varphi)$ of an aspherical lens, where $\varphi$ and $r(\varphi)$ are angle and radius-vector of the point of the polar coordinate system, so as to exclude its spherical aberration. I.e. all falling normally on the lens rays (shown in Figure 1. in red) were collected at a single point (the focus) without deviation.

To do this, we draw up a differential equation of the form $f(r,r',\varphi)=0$ whose solution gives the desired curve. The refractive index of the lens assumed to be $n$, refractive index of air is assumed to be 1.

Figure 1 shows the profile of the lens with additional constructs required to solve the problem. For point B the tangent, the perpendicular and the polar radius are drawn. The coordinate axes on all figures can be marked with any spatial quantities (from millimeters to meters) depending on the size of the lens.

Figure 1 shows that the focal length $F$

$$F = r\cos\varphi + r\sin\varphi \cdot \cot\delta \tag{6}$$

From triangle ABC: $\alpha + 180 - \gamma + \delta = 180$, thence

$$\delta = \gamma - \alpha \tag{7}$$

Using the well-known trigonometric formula for the cotangent of the difference of two angles

$$\cot\delta = \cot(\gamma - \alpha) = \frac{1 + \tan\gamma \cdot \tan\alpha}{\tan\gamma - \tan\alpha} \tag{8}$$

From (8) should be excluded $\tan\gamma$ and $\tan\alpha$, expressing them through $\varphi$.

In view of law of refraction

$$n\sin\alpha = \sin\gamma \tag{9}$$

$$\tan\gamma = \frac{n\sin\alpha}{\cos\gamma} \tag{10}$$

and

$$1 - n^2\sin^2\alpha = \cos^2\gamma \tag{11}$$

In view of (11)

$$\tan\gamma = \frac{n\sin\alpha}{\sqrt{1 - n^2\sin^2\alpha}} \tag{12}$$

dividing the numerator and denominator (12) on $\sin\alpha$ and taking into account the trigonometric ratio

$$\sin^2\alpha = \frac{1}{1 + \cot^2\alpha} \tag{13}$$

obtain

$$\tan\gamma = \frac{n}{\sqrt{1 + \cot^2\alpha - n^2}} \tag{14}$$

then

$$\cot\delta = \frac{\sqrt{1 + \cot^2\alpha - n^2} + n \cdot \tan\alpha}{n - \tan\alpha\sqrt{1 + \cot^2\alpha - n^2}} \tag{15}$$

Now it is necessary to exclude $\alpha$ in (15). From the mathematical analysis for the function given in polar coordinates is known that [6]

$$r' = r\tan(\varphi - \alpha) \tag{16}$$

In view of (16), and the formula for the difference of two angles tangent

$$\tan\alpha = \frac{r\tan\varphi - r'}{r'\tan\varphi + r} \tag{17}$$

Substituting (17) into (15) and (15) to (6) the desired differential equation is obtained

$$F = r\cos\varphi + r\sin\varphi \cdot \frac{\sqrt{1 - n^2 + \left(\frac{r'\tan\varphi + r}{r\tan\varphi - r'}\right)^2} + n \cdot \frac{r\tan\varphi - r'}{r'\tan\varphi + r}}{n - \frac{r\tan\varphi - r'}{r'\tan\varphi + r}\sqrt{1 - n^2 + \left(\frac{r'\tan\varphi + r}{r\tan\varphi - r'}\right)^2}} \tag{18}$$

For small angles of incidence $\alpha$, i.e. for long-focus lenses equation (18) can be simplified by taking in (15)

$$\cot^2\alpha \gg 1 - n^2 \tag{19}$$

and

$$\cot\alpha \gg n\tan\alpha \tag{20}$$

obtain

$$\cot\delta = \frac{\cot\alpha}{n - 1} \tag{21}$$

then (18) can be written as follows

$$F = r\cos\varphi + r\sin\varphi \cdot \frac{r'\tan\varphi + r}{(r\tan\varphi - r')(n - 1)} \tag{22}$$

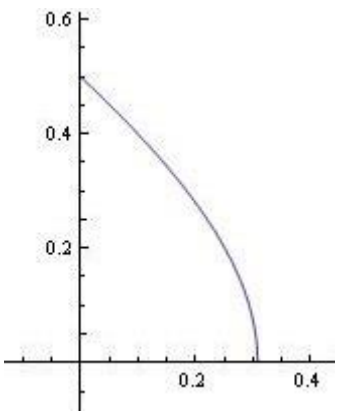

**Figure 2.** Meridian profile of the aspheric lens with short focus.

Differential equations (18) and (22) were solved using the Wolfram Mathematica package. The solution of equation (18) for *n=1,2*, *F=2* with the boundary condition $r(\pi/2) = 0,5$ (spatial units can be from millimeters to meters, depending on the size of the lens) is shown in Figure 2. Unfortunately Mathematica package failed to solve this equation in the range from $-\pi/2$ to $\pi/2$.

The solution of equation (23) for *n=1,2*, *F=7* with the boundary condition $r(0) = 0,15$ is shown in Figure 3.

## 3. Conclusions

A method for finding the meridian profile of an aspherical lens was proposed by compiling and numerically solving the differential equation. The problem was solved by compiling the differential equation of an aspherical lens and the meridian profile of the aspherical lens was obtained. The conditions to simplify the equation were found for long-focus lenses. Equations (18) can describe aspheric lenses with any focuses, but more convenient to use it for short-focused lenses. Simplified differential equation (22), that written for small angles of incidence on lens surface well describes long-focus aspheric lenses.

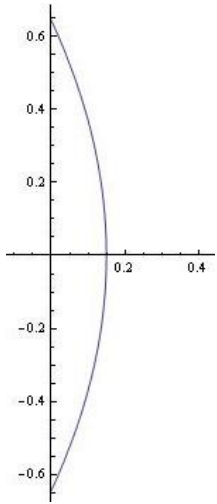

**Figure 3.** Meridian profile of the aspheric lens with long focus